\documentclass[preprint]{aastex}

\shorttitle{First Blue Stragglers in the Bulge}
\shortauthors{Clarkson et al.}

\begin{document}

\title{The First Detection of Blue Straggler Stars in the Milky Way Bulge}
\author{William I. Clarkson\altaffilmark{1,2}, Kailash C. Sahu\altaffilmark{3}, Jay Anderson\altaffilmark{3}, R. Michael Rich\altaffilmark{2}, T. Ed. Smith\altaffilmark{3}, Thomas M. Brown\altaffilmark{3}, Howard E. Bond\altaffilmark{3}, Mario Livio\altaffilmark{3}, Dante Minniti\altaffilmark{4,5}, Alvio Renzini\altaffilmark{6} and Manuela Zoccali\altaffilmark{4}}

\affil{\altaffilmark{1}Present address: Department of Astronomy, Indiana University, Bloomington, 727 East 3rd Street, Swain West 319, Bloomington, IN 47405-7105, USA}
\affil{\altaffilmark{2} Division of Astronomy and Astrophysics, University of California, Los Angeles, 430 Portola Plaza, Box 951547, Los Angeles, CA 90095-1547}
\affil{\altaffilmark{3}Space Telescope Science Institute, 3700 San Martin Drive, Baltimore, MD 21218, USA}
\affil{\altaffilmark{4}Departamento de Astronomia y Astrofisica, Pontifica Universidad Catolica de Chile, Vicuna Mackenna 4860, 7820436, Santiago, Chile}
\affil{\altaffilmark{5} Vatican Observatory, Vatican City State V-00120, Italy}
\affil{\altaffilmark{6} INAF - Osservatorio di Padova, Vicolo dell'Osservatorio 5, 35122, Padova, Italy}
\email{wiclarks@indiana.edu}

\begin{abstract}

We report the first detection of Blue Straggler Stars (BSS) in the
bulge of the Milky Way galaxy. Proper motions from extensive
space-based observations along a single sight-line allow us to
separate a sufficiently clean and well-characterized bulge sample that
we are able to detect a small population of bulge objects in the
region of the color-magnitude diagram commonly occupied young objects
and blue strgglers.  However, variability measurements of these
objects clearly establish that a fraction of them are blue stragglers.
Out of the 42 objects found in this region of the color-magnitude
diagram, we estimate that at least 18 are genuine BSS.  We normalize
the BSS population by our estimate of the number of horizontal branch
stars in the bulge in order to compare the bulge to other stellar
systems. The BSS fraction is clearly discrepant from that found in
stellar clusters.  The blue straggler population of dwarf spheroidals
remains a subject of debate; some authors claim an anticorrelation
between the normalised blue straggler fraction and integrated
light. If this trend is real, then the bulge may extend it by three
orders of magnitude in mass.  Conversely, we find that the {\it
  genuinely} young ($<$~5Gy) population in the bulge, must be at most
3.4\% under the most conservative scenario for the BSS population.
\end{abstract}

\keywords{Galaxy:bulge – Galaxy:disk - Galaxy:evolution}

\section{Introduction}

Blue Straggler Stars (hereafter BSS) are thought to be old,
hydrogen-burning stars, made hotter and more luminous by accretion and
now residing in a region of the color-magnitude diagram (CMD) normally
occupied by much younger stars
\citep[e.g.][]{bond71,livio93,stryker93}. They have been detected in
examples of nearly every stellar system - open clusters
\citep[e.g.][]{ahumada07}, globular clusters
\citep[e.g.][]{piotto04,lanzoni07}, dwarf spheroidal (dSph) galaxies
\citep{momany07, mapelli07} and the Milky Way stellar halo
\citep{ps00,carney01}. However, while BSS in the bulge have long been
suspected \citep[e.g.][]{morrison93}, their existence in the bulge as
a population has never been proven due to strong contamination from
disk stars that fill the same region in the CMD.\footnote{The
  hypervelocity B-type star (HVS) HE 0437-5439 is most likely a BSS
  from the inner milky way, although its BSS nature may result from
  the ejection-encounter \citep{brownw10}.}

By the standards of modern space-baced astrometry, disk and bulge
stars have large relative motions ($\sim5$~mas yr$^{-1}$) and thus a
reasonably pure sample of bulge stars can be isolated
kinematically. Prior use of the Hubble Space Telescope (hereafter HST)
to perform this separation indicated that the population of bulge blue
stragglers must be small ($\le$~few percent), but uncertainty in the
kinematic contamination (and thus foreground disk objects mimicking
BSS), combined with the absence of corroborating evidence like
variability, prevented constraint on the BSS population size
\citep{kr02}. As a result, BSS have never unambiguously been detected
in the Milky Way bulge.

Beyond the intrinsic interest of BSS populations in a predominantly
old stellar population, detection of BSS in the Milky Way bulge would
be important for galactic structure formation. Most spiral galaxies
show flattened ``Boxy/Peanut'' bulges (see \citealt{ath05} and
\citealt{bureau_ath05} for a clear discussion) or a small ``Disky
bulge'' accompanied by ongoing star formation \citep{kk04}. Such a
bar-driven arrangement of the stars evolves over a timescale of
several Gy, and can be transient with multiple generations of bar
\citep[e.g.][]{combes09}. However, chemical evidence suggests the
stars themselves mostly formed early and rapidly
\citep[e.g.][]{ballero07, mcwilliam10}. While the majority view now is
that most stars in the Boxy/Peanut bulge of the Milky Way (which we
call here ``the bulge'') formed $\sim$10 Gy ago \citep[e.g.][]{zocc03,
  freeman08}, the degree to which this star formation was extended
over time, is presently only somewhat weakly constrained
\citep{kr02,zocc08}. Thus, better estimates of the young stellar
population of the bulge, and thus its evolution, are a natural
dividend of constraints on the BSS fraction.

We report our use of a combination of proper motion and variability
information to set this constraint. The observations are detailed in
Section \ref{s_obs}. Section \ref{s_samplesel} explains the selection
of candidate BSS and the analysis used to estimate the size of the
true BSS population. Finally in Sections \ref{s_discuss} \&
\ref{s_conclude} we discuss the BSS population in the context of
stellar populations in general and the formation of the bulge of the
Milky Way in particular.

\section{Observations}\label{s_obs}

All motions on which we report here were 
obtained from two-epoch imaging of the SWEEPS (Sagittarius Window
Eclipsing Extrasolar Planet Search) field (Sahu et al. 2006) towards
the bulge (at $l,b=2.65\degr,-1.25\degr$) using the Wide Field Channel
of the Advanced Camera for Surveys on HST, which has pixel scale 49.7
mas pix$^{-1}$ \citep{vandermarel07}. The sight-line passes through
the bulge just beneath the galactic mid-plane, passing within
$\sim$350pc of the galactic center
(but always exterior to the nuclear cluster, which shows ongoing
  star formation; \citealt{launhardt02}).  Full details on the analysis of
these extensive HST datasets can be found in \citet[][hereafter Paper
  I]{c08}, Sahu et al. (2006) and \citet{gillil08}; we report here the
aspects relevant to BSS detection and characterization. 

\subsection{Astrometry and Photometry} \label{s_ast}

The measurements used are: (i) proper motions down to $F814W = 24$~in
the SWEEPS field as measured from astrometry in the two epochs; (ii)
absolute photometry in $F606W$~\& $F814W$, and (iii) time-series
photometry of the stars from the continuous seven-day SWEEPS campaign,
with gaps only due to Earth occultation.

The techniques used to measure proper motions from the long (339 sec)
and short (20 sec) exposures were detailed in Paper I. For objects
saturated in the long exposures, motions and errors using the short
exposures are substituted. The stellar time-series and absolute
photometry were taken from the work of \citet{sahu06}, in which full
details can be found. Difference image analysis techniques
\citep[][and references therein]{gillil08} were used to produce the
time-series; the noise in the result is very close to Poisson for the
majority of objects, and perhaps 30\% higher than Poisson for
saturated objects. Absolute magnitudes were produced from psf fitting
to the image model against which variability was measured (Sahu et
al. 2006).

\section{Sample selection} \label{s_samplesel}  

The bulge population was extracted using cuts on motion and
measurement quality, and the BSS candidates evaluated from this
sample (Section \ref{s_selreg}), along with contamination by disk and
halo stars that fall into the sample (Section
\ref{s_kcontam}). Variability information was used to set limits on
the size of the young ($\la 5$~Gy) population among bulge BSS
candidates (Section \ref{s_var}), corrected for periodicity detection
incompleteness (Section \ref{s_complete}). This allowed the size of
the true BSS population in the bulge to be bounded (Section
\ref{s_bounds}).

\subsection{Separation of bulge from field} \label{s_bulgesep}\label{s_selreg}

The streaming of disk stars in front of the bulge in galactic
longitude was used to isolate the bulge population. Proper motions were
measurable for 188,367 of the 246,793 stars with photometry in the
field of view. Stars were selected for further study with proper
motion error $\sigma_l \le 0.3$~mas yr$^{-1}$~and $\sigma_b \le
0.45$~mas yr$^{-1}$, and with crowding parameter $q < 0.1$~(see
Paper I for definition of the crowding parameter). This latter limit
is slightly more restrictive than used in Paper I, and results in a
well-measured astrometric sample of 57,384 objects down to $F814W
\simeq 23$. Discarding objects saturated in the short exposures then
leaves 57,265 well-measured, unsaturated objects. To set the proper
motion cut for bulge stars, the distribution of latitudinal proper
motion $\mu_l$~was measured for stars above the main sequence turn-off
(hereafter MSTO), where the disk+BSS and bulge populations trace
separate loci in the CMD. Using a proper motion cut $\mu_l < -2.0$~mas
yr$^{-1}$, rejects all but 0.19\% of disk stars, while admitting 26\%
of bulge objects. With error cut at 0.3~mas yr$^{-1}$~and $\mu_l <
-2.0$~mas yr$^{-1}$, galactic-longitudinal motions must be detected to
at least 6$\sigma$~for inclusion in the cleaned bulge
catalogue. Applied to the full error-cleaned sample with astrometry,
the motion cut produced a population of 12,762 likely-bulge objects.

The BSS selection region was set with the constraints that (i). the
bright end should not impinge on the region in which blue horizontal
branch objects are expected, and (ii). the faint and red ends should
not greatly overlap with the main population of the bulge near the
MSTO (see Figure 1 for the selection region adopted in the $F814W$,
$F606W-F814W$ CMD). Within this selection-region, 42 objects were
found in the kinematically-selected bulge sample.

\subsection{Kinematic and photometric contaminants}\label{s_kcontam}

The number of disk objects amongst our BSS candidate sample was
estimated directly from the distribution of longitudinal proper motion
$\mu_l$. Beginning with the 57,265 objects that are astrometrically
well-measured, the longitudinal proper motion distribution is assessed
within the post-MSTO bulge sample as shown in Figure
\ref{fig_kcontam}~(top panel). The CMD selection-region for this
post-MSTO bulge sample was constructed by inverting the BSS selection
box in color about its reddest point, while keeping the same magnitude
range as the BSS sample. We denote the set of longitudinal proper
motions of these post-MSTO bulge objects as $\mu_{l,bulge}$. It is
clear from Figure 2 that the contribution of any disk objects to this
population is very small. This population was fit with a single
gaussian component.

The longitudinal proper motion distribution within the BSS selection
region we denote by $\mu_{l,BSS}$, and was fit with a model consisting
of two gaussian components, one ($\mu_{l,1}$)~representing disk
objects, one ($\mu_{l,2}$)~representing bulge objects (Figure
\ref{fig_kcontam}, middle panel). The centroid and width of
$\mu_{l,2}$~were frozen to the values fit from $\mu_{l,bulge}$;
because the BSS and post-MSTO bulge samples cover an identical
magnitude range, to first order they should suffer similar
instrumental and incompleteness effects. The expected number of disk
objects within the BSS selection-box (denoted $N_d$)~was evaluated
from the parameters of the best fit to $\mu_{l,BSS}$~with these
constraints (the best-fit to $\mu_{l,BSS}$~shows $\chi^2_{\nu} =
24.2/26$).

We performed a Monte-Carlo simulation where a large number of proper
motion datasets drawn from the best-fit model were generated, and the
distribution assessed of recovered values of both the number of disk
objects in the BSS region $N_d$ and the number that would pass our
kinematic cut for bulge objects ($N_{d,<-2.0}$). The results are shown
on the bottom panel of Figure \ref{fig_kcontam}. The recovered
$N_{d,<-2.0}$~is $0.7\pm0.73$~objects, but the distribution is rather
asymmetric - while the most common recovered $N_{d,<-2.0}$~suggests
fewer than one contaminant, in some cases up to four contaminants are
measured. We therefore adopt 0-4 as the ranges on the contribution
from disk objects to our kinematically cleaned bulge sample.

Of the $\sim 6$~estimated halo contaminants (Paper I), of order 0-2
objects would fall into the BSS selection region. So, of 42 objects,
we expect $0-6$~objects to have passed our kinematic cuts that do not
belong to the bulge.

We investigated by simulation the contamination in our BSS region due
to the main bulge population near the MSTO, which might be expected to
put outliers in the BSS region.  Synthetic bulge populations were
constructed with age range $7$-$13$~Gy \citep{zocc03} and metallicity
distribution approximating the bulge \citep{santiago06,zocc08}. Age
and metallicity samples were converted to predicted instrumental
magnitudes using the isochrones described in \citet{brown05}. The
resulting population was then perturbed by the bulge distance
distribution and with a binary population estimated using a variety of
binary mass-ratio distributions and binary fractions
\citep{soderhjelm07} to produce simulated bulge-selected populations
in the CMD. The resulting number of objects from the main bulge
population that happen to fall in the BSS region in the CMD, is
$5\pm2$, over $10^5$ trials.

We therefore estimated the total contaminant population at (5-13)/42
objects, leaving (29-37)/42 possible bulge BSS.

\subsection{Photometric Variability}\label{s_var}

BSS may form by a variety of mechanisms \citep[e.g.][]{abt85,
  livio93}, including mass transfer in a binary system
\citep{mccrea64}. This can yield BSS that are presently in binaries
\citep[e.g.][]{andronov06,chenhan09} with observable radial-velocity
periodicities \citep[e.g.][]{carney01,mathieu09}. For periods $\la$10
days, photometric variations caused by tidal deformation of one or
both members may be observed, in the extreme producing a W UMa-type
contact binary \citep{mateo93,mateo96}.

The variability information from the 2004 epoch (Section \ref{s_obs})
allows the tendency for W UMa variability among the BSS candidates to
be examined.  In a manner similar to Section \ref{s_bulgesep}, a
population composed of mostly disk objects was isolated, this time
using longitudinal proper motion $\mu_l \ge +3.0$~mas yr$^{-1}$~and
again requiring 6$\sigma$~motion detection. This test sample indicates
the fraction of W UMa-like variables among a young population observed
at the same signal to noise range as the bulge objects we wish to
probe. Comparison of the W UMa occurrence rate between the mostly-disk
and mostly-bulge samples then provides an estimate of the fraction of
genuine BSS among the BSS candidates.

To avoid confusion with pulsators, systems were {\it not} counted
among the W UMa if the dominant period detected $P$~was shorter than
$0.5$d, or if significant phase or frequency variation was present
along the observation interval \citep{gillil08}.  Note that we do not
eliminate possible pulsators from our BSS candidate table, as BSS can
also show photometric pulsations \citep{mateo93}.  Each variable
lightcurve was visually examined to determine if its morphology
resembles that of W Uma objects or close eclipsing binaries found in
other systems \citep[e.g.][]{mateo93,albrow01}. W UMa's were much more
populous amongst the bulge-selected population: of the 42
bulge-selected objects, four show W UMa-type variability (one
additional object is apparently a long-period pulsator; Figure
\ref{fig_sinus}). Of this four, one object only has a single eclipse
event recorded and awaits confirmation. Conversely, {\it none} of the
81 disk-selected objects show W UMa photometric variability.

\subsection{Periodicity completeness correction}\label{s_complete}

The completeness of the Lomb-Scargle (hereafter LS) variability search
to W UMa variables was estimated through simulation. Real lightcurves
with no significant periodicity detection (peak LS power $<$5) were
injected with W UMa-type lightcurves \citep[following the description
  of][]{rucinski93} under a log-uniform period distribution. Synthetic
W UMa systems were denoted ``detected'' if the peak in the
Lomb-Scargle periodogram was within 10\% of the input period, the
signal was present in the lightcurves in both filters, and showed LS
peak power greater than 13 (the 99\% significance level, defined as
the peak LS power exceeding that returned by 99\% of trials with no
periodicity injected). The process was repeated for input variation
amplitudes $\Delta$mag/mag = 0.1, 0.05, 0.02, 0.01, 0.005, 0.002,
0.001 and period bins 0.4, 0.8, 1.7, 3.4 and 7.0d (Figure
\ref{fig_sinus}). The dependence of the detection completeness on
injected period is rather weak but the dependence on amplitude is
clear; at least $88\%$ of systems with 2\% fractional amplitude
variation were detected, and 70\% at the 0.5\% amplitude level.

To map signal amplitude to system inclinations, synthetic W UMa
lightcurves were simulated using the \citet{wd71} code as implemented
in the Nightfall
package\footnote{http:/\//\/www.hs.uni-hamburg.de/\/DE/\/Ins/\/Per/\/Wichmann/\/Nightfall.html}
indicating that at most $\sim$70\% of W UMa type systems would be
detectable down to binary inclination $\sim$45$\degr$, so we are
sensitive to perhaps 35\% of the true W UMa population. This implies
that the 3-4 detected W UMa trace an underlying population of 9-11
systems with W UMa-like system parameters, so at least $9$-$11$~of the
42 BSS candidate objects have W UMa system parameters, and are therefore
not young.

If, as our measurements suggest, 9-11 of the 42 BSS candidates are
present-day W UMa objects, we can ask what fraction of the BSS
candidates are in binaries of {\it any}~orbital period. The period
distribution for bulge BSS {\it presently} in binaries is unknown; we
use observations of the solar neighbourhood and open clusters as
proxies for the bulge population.

\citet{carney01} report spectroscopic periods for 6/10 high-proper
motion blue metal-poor (hereafter BMP) stars in the solar
neighborhood, isolating BSS candidates by virtue of de-reddened colors
bluer than the MSTO of globular clusters of comparable metallicity
\citep{carney94}. That study also reports a re-analysis of the radial
velocities of the BMP sample of \citet{ps00}, finding ten
spectroscopic BMP orbits for a grand total of sixteen spectroscopic
orbits among 6 BSS candidates and 10 further BMP that may be BSS. This
yields 2/16 objects with $\log_{10}(P_{orb}) < 1$~and 14/16 with $1.0
\le \log_{10}(P) < 3.2$~(with one possible very long-period binary not
included in this sample). Thus, in the solar neighborhood perhaps 1/8
of the total binary BSS population resides in short-period binaries, with
$\log_{10}(P_{orb}) < 1.0$.

Turning to the open clusters, \citet{mathieu09}~find a high binary
fraction and broadly similar period distribution amongst BSS
candidates in the $\sim 7$~Gy-old open cluster NGC 188. There, 16/21
BSS candidates are in binaries. Of these sixteen objects, two show
$\log_{10}(P) < 1.0$~while the remaining 14 all show $1.0 \le
\log_{10}(P) < 3.5$. For the $\sim 5$~Gy-old open cluster NGC 2682
(M67), \citet{latham07} reports two BSS in binaries with
$\log_{10}(P_{orb} < 1)$~and five with $1.0 \le \log_{10}(P_{orb}) <
3.1$ \citep[Figure 3 of][]{peretsfab09}. Thus, in open clusters the
short-period binaries may make up between about 1/8 and 2/7 of the
total population of BSS in binaries.

The period distribution for main-sequence binaries of several types is
collated in \citet{kroupa95}. For $P_{orb} \la 3.5$~the number of
binaries increases with $P_{orb}$~at least as steeply as a log-uniform
distribution \citep{kroupa95}. If BSS in binaries in the bulge follow
a similar distribution, then binary BSS with $\log_{10}(P) < 1$~would
make up at most about 1/3 of the total population with $\log_{10}(P) <
3.0$.

Taken together, these estimates suggest that, if all the W UMa are
indeed BSS presently in binaries, then BSS in binaries could {\it by
  themselves} account for {\it all} the 29-37 non-contaminant objects
among our candidate BSS.

\subsection{Bounds on the true BSS population in the bulge}\label{s_bounds}

As none of the kinematically-selected disk objects show W UMa
characteristics, the W UMa stars in the BSS region are unlikely to be
disk stars. Furthermore, W UMa variability is expected to be a natural
evolutionary state of some of the blue stragglers. Thus, the detection
of W UMa variability in the BSS region of the bulge CMD certifies
these stars as true bulge blue stragglers. Hence, we conclude that the
$9-11$~W UMa objects are in fact bulge BSS, from a possible bulge BSS
population of $29$-$37$~(Section \ref{s_kcontam}). This leaves
$18$-$28$~bulge objects whose nature remains undetermined. It is
unlikely that W UMa variables make up the entire BSS population in the
bulge. Based on arguments in Section \ref{s_complete}, we consider the
BSS population with binary periods $\log_{10}(P) < 1$~to comprise {\it
  at most} half of the population of BSS with any orbital period. We
therefore consider two extreme scenarios: 1. {\it Optimistic}: all of
the remainder are bulge BSS – thus there are no genuinely young bulge
objects in our sample, leaving $29-37$~BSS, and 2. {\it Conservative}:
our W UMa objects make up half of the BSS among our sample, suggesting
18-22 genuine BSS and $7-19$~possibly young bulge objects.

In addition, a significant fraction of the bulge BSS population may
exist with no binary companion or in very wide binaries, as indicated
by population studies \citep[e.g.][]{andronov06, chenhan09} and
observations of clusters and the solar neighbourhood
\citep[e.g.][]{ferraro06,carney01}. Therefore the true BSS population
in the bulge probably lies between the ``conservative'' and
``optimistic'' scenarios outlined above.

\section{Discussion}\label{s_discuss}

\subsection{Normalization of $N_{BSS}$}\label{s_normal}

As probes of binary evolution, BSS indirectly probe the star formation
history of stellar populations in their own right. To compare the
bulge BSS population with that of other stellar systems, we normalize
by the number of horizontal branch objects in the system; $S_{BSS} =
N_{BSS}/N_{HB}$. We adopt $N_{HB} = N_{BHB} + N_{RHB} + N_{RC}$, where
the three terms denote the Blue and Red horizontal branch stars, and
the Red Clump stars respectively, and thus cover the full metallicity
range of post-MS core He-burning objects.  This is the convention used
for the dwarf spheroidals \citep[e.g.][although some authors choose
  not to include $N_{RC}$~explicitly if none are found in their
  sample]{carrera02, mapelli07, momany07}; open clusters
\citep{demarchi06} and the globular clusters \citep[][though here the
  definition of $N_{HB}$~is not explicitly stated]{piotto04}. The
exception is the stellar halo, where $N_{HB}=N_{BHB}$~is adopted
because only BHB can be distinguished kinematically and
photometrically from the local disk population \citep{ps91a}.

The kinematic cuts leave too sparse a population to accurately
disentangle the RC from the underlying Red Giant Branch (RGB). We
therefore estimate $N_{RC}$~from the $F606W$~histogram along the RGB
only after selection by astrometric error and crowding parameter, and
scale the result to the bulge population passing the full set of cuts
(Section \ref{s_bulgesep}).  Figure \ref{fig_nhb} shows the selection;
the marginally-detected RGB Bump (hereafter RGBB) is masked out as
well as the RC before fitting the continuum due to the RGB; Gaussian
components fit to the RC and RGBB then yield the number of objects in
each population over the RGB continuum. Bootstrap monte carlo sampling
before producing the histogram is used to estimate the fitting
uncertainty for a given binning scheme. We find $N_{RC}= 180 \pm
53$~while the RGBB is detected at more marginal significance;
$N_{RGBB} = 41 \pm 23$.\footnote{\citet{nataf11}~recently found an
  anomalously small RGBB population compared to the RC, with
  $N_{RGBB}/N_{RC}$=$(12.7 \pm 2)\%$~among a large number of
  sight-lines across the bulge. We do not confirm this finding
  \citep[compare our Figure 4 with Figure 1 of][]{nataf11}, although
  with $N_{RGBB}/N_{RC}$=$(23 \pm 15)\%$ our statistics are too poor
  to falsify their claim of He-enrichment. In addition, sampling as we
  do towards the center of the ``X''~structure in the bulge
  \citep{nataf10,McW_Zocc_10}, an intrinsic difference in
  $N_{RGBB}/N_{RC}$~may be expected between our sight line and the
  range reported by \citet{nataf11}.} Our estimate of $N_{RC}= 180 \pm
53$~scales to the 26\% of bulge objects surviving the full kinematic
cuts, as $47 \pm 14$~objects. Even in the short exposures, roughly
15\% of this population would saturate and thus not enter the bulge
sample (Figure \ref{fig_nhb}), so that we would expect $40 \pm 12$~RC
objects to fall within the bulge sample. This figure includes the RHB objects. 

\begin{figure}
\epsscale{0.60}
\plotone{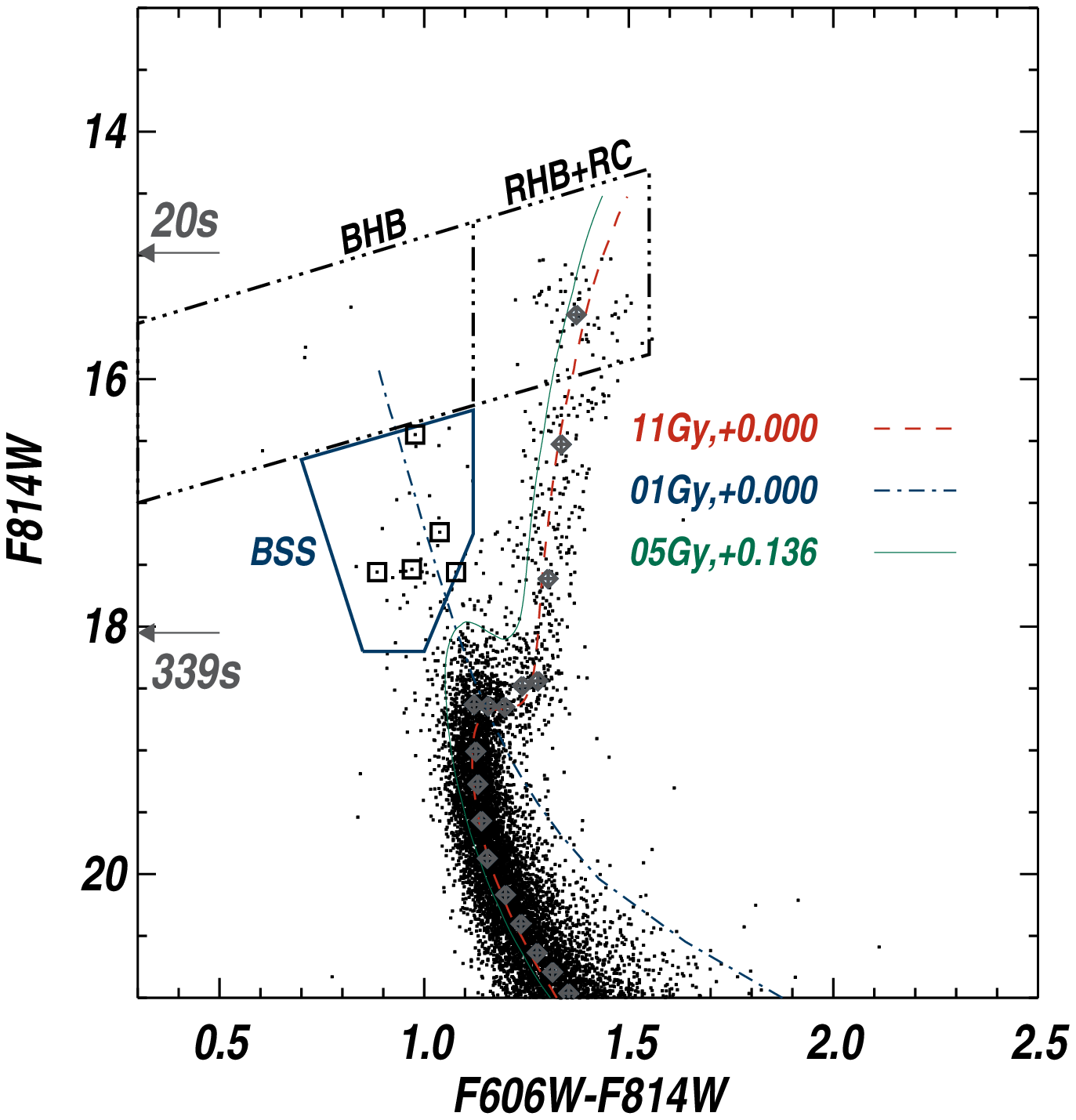}
\caption{Color-Magnitude Diagram (CMD) of the kinematically selected extreme-bulge population. Isochrones are shown for a young foreground population (blue dash-dot), old bulge population (dashed) and an intermediate-age, slightly metal-poor bulge population (solid). Diamonds show the median bulge population. Ages are indicated, as are the adopted $[Fe/H]$~values. Selection regions for BSS and horizontal branch objects are indicated. Saturation limits in the short and long exposures are marked, and the five BSS candidates exhibiting close binary variable (W UMa) lightcurves are indicated. See Section \ref{s_selreg}.}
\label{fig_cmd}
\end{figure}

\begin{figure}
\epsscale{1.20}
\plottwo{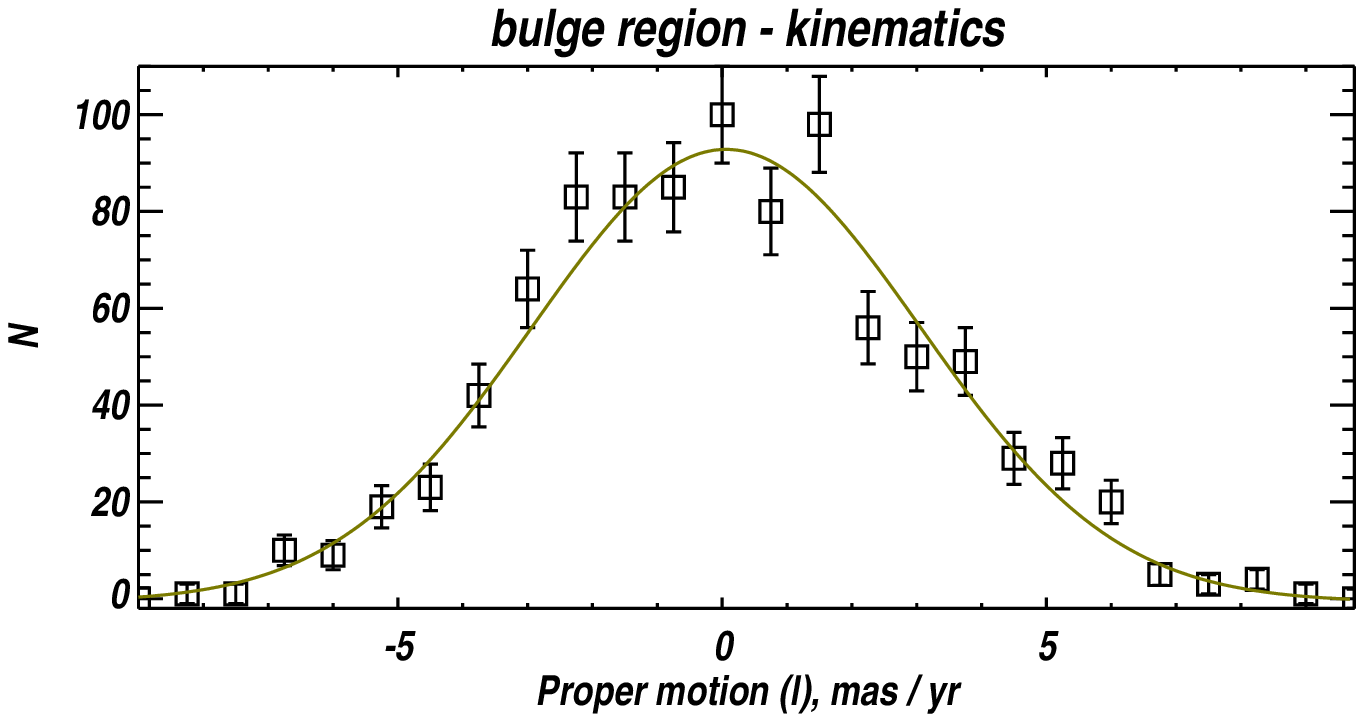}{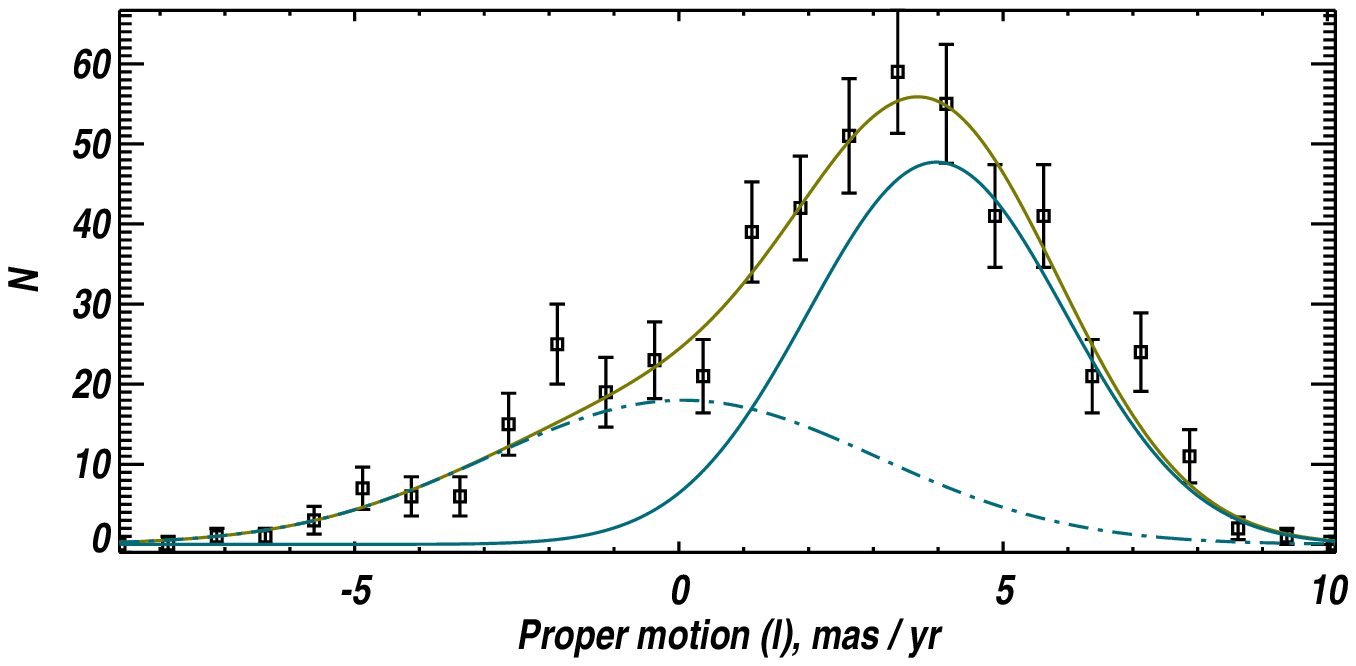}
\epsscale{0.60}
\plotone{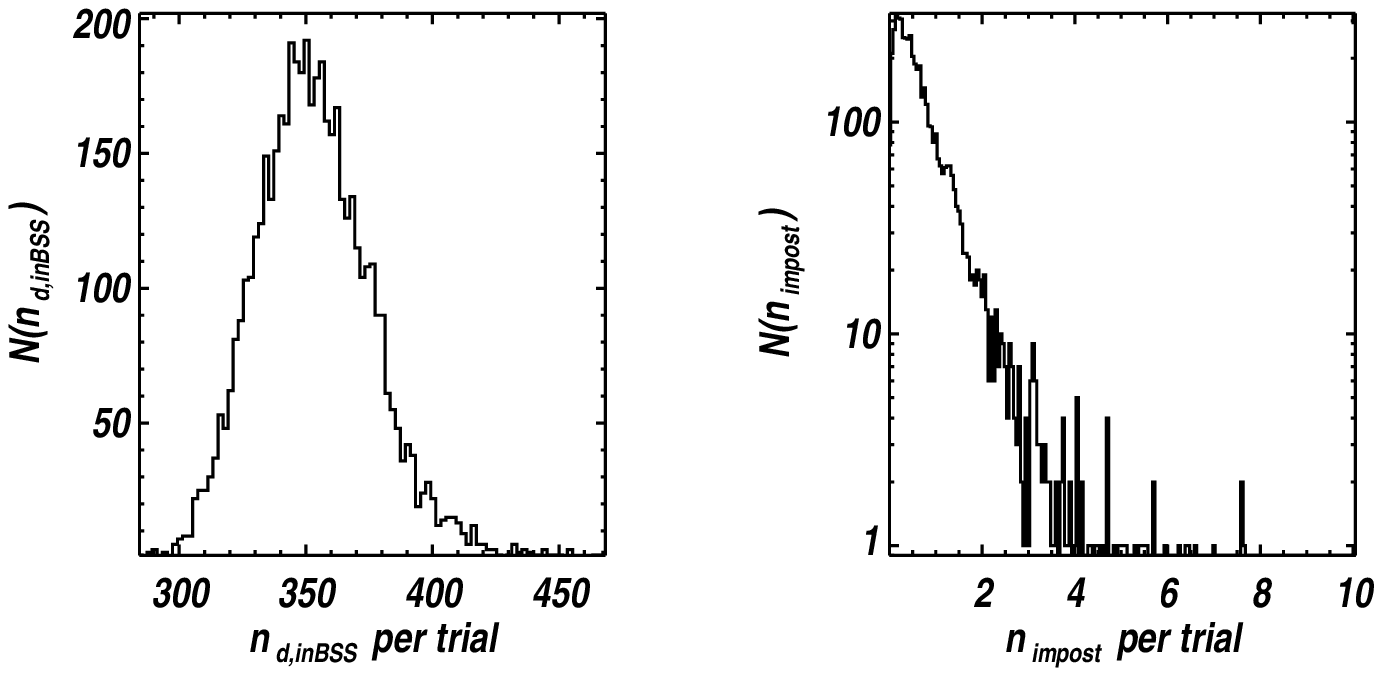}
\caption{Estimation of the kinematic contamination in the BSS CMD-selection region (Figure \ref{fig_cmd}), due to disk objects. {\it Top:} single-component gaussian fit to the distribution of longitudinal proper motion for post-MSTO bulge objects ($\mu_{l,bulge}$), which we use to calibrate the bulge component in the BSS region. {\it Middle:} two-component gaussian model fit to the longitudinal proper motion distribution within the BSS region of the CMD ($\mu_{l,BSS}$). {\it Bottom-Left:} distribution of recovered number of disk objects in the BSS region under 5,000 Monte Carlo trials. {\it Bottom Right:} Distribution of recovered number of disk objects that would pass the kinematic cut $\mu_l < -2.0$~mas yr$^{-1}$. See Section \ref{s_kcontam}.}
\label{fig_kcontam}
\end{figure}

\begin{figure}
\epsscale{0.8}  
\plotone{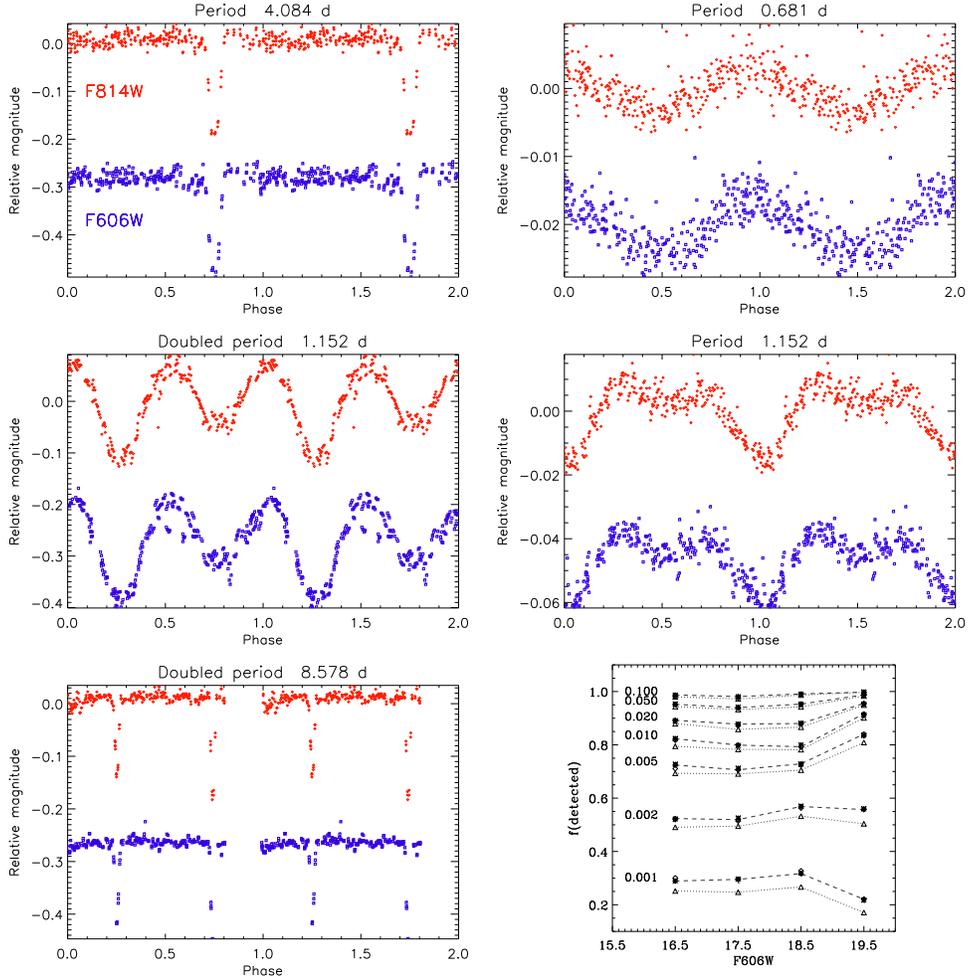}
\caption{Folded lightcurves for the five objects showing sinusoidal or
  eclipse-type variability with a period longer than 0.5 days. The
  object at the top-right is probably a long-period pulsator. The
  objects at middle-left and bottom-left were detected at half the
  likely true period, as suggested by differing eclipse depths half a
  period apart. {Bottom Right:} Completeness to W UMa variability from
  the HST dataset. {\it Left:} fraction of injected W UMa lightcurves
  recovered as a function of magnitude and amplitude. Injected
  variation amplitudes $\Delta V/V$~are labeled next to each set of
  curves. For each set of curves, from top to bottom, symbols
  represent periods 0.4, 0.8, 1.7, 3.4 and 7.0d respectively; in
  practice, all the limits for a given amplitude overlap except for
  the 7d set.}
\label{fig_sinus}
\end{figure}

\begin{figure}
\plotone{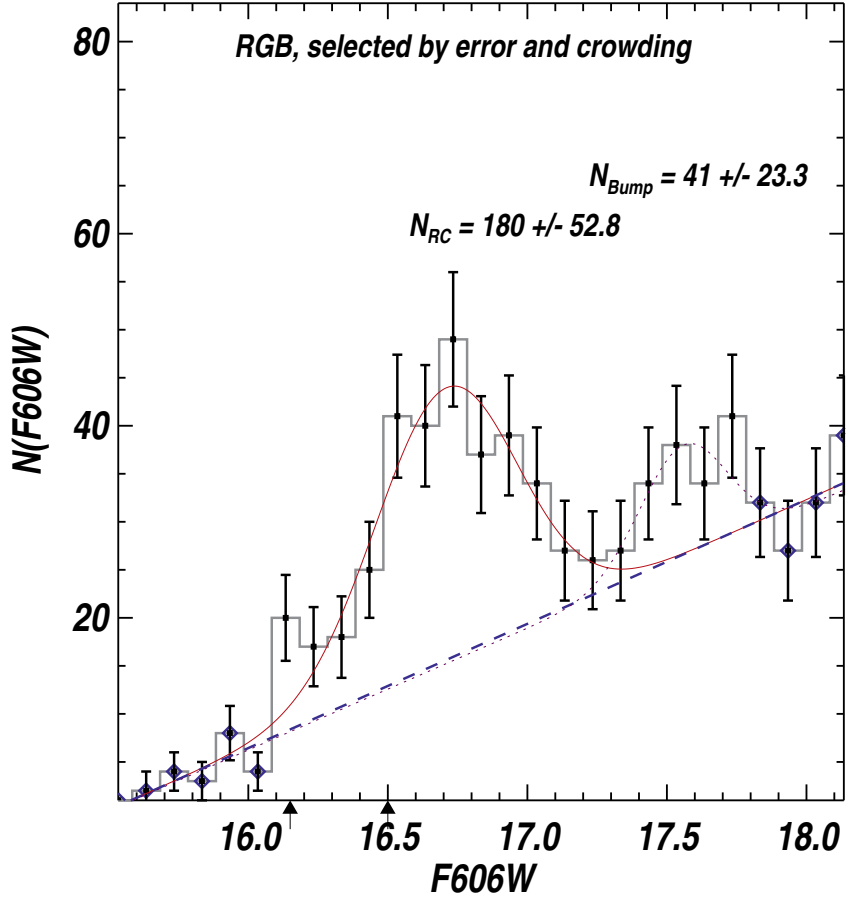}
\caption{Estimation of the red clump population $N_{RC}$~for the bulge. We use the sample selected for astrometric error and crowding, and use the population within the bulge RGB rather than cutting by proper motion to afford sufficient statistics to constrain $N_{RC}$~accurately. Diamonds show bins selected for the fit to the underlying RGB population (dashed line), subtracted before estimating $N_{RC}$. The fits to the RC population (solid line) and RGB bump (dotted line) are indicated, as are the population sizes in the RC and RGB bump (the latter is marginally detected). Triangles indicate astrometric saturation limits for the $RC$ at the blue and red ends of the selection box (Figure 1). See Section \ref{s_normal}}
\label{fig_nhb}
\end{figure}

The contribution from BHB is difficult to estimate for the bulge, but
is clearly rather small - it vanishes against the much larger
foreground disk population without kinematic selection. Unlike the
RC+RHB selection region, saturation of the astrometric exposures is
not an issue for the BHB selection region we adopt (Figure
\ref{fig_cmd}). Altogether, four objects are found in the BHB + RHB
selection region; of order one object may be a kinematic
contaminant. We therefore consider $(3-4 \pm 1\sigma_{\rm Poisson}) =
2-6$~as the limits on $N_{BHB}$.

Thus we estimate the total $N_{HB}=$30-58 objects among the same
sample from which the BSS were estimated. Our estimate of $18 \le
N_{BSS} \le 37$, then leads to $0.31 \le (S_{BSS} = N_{BSS}/N_{HB}) \le
1.23$.

\subsection{Comparison to other systems}\label{s_compare}

For the halo, \citet{ps00}~estimate $N_{BSS}/N_{BHB} = 5$, though their
selection criteria appear to admit a population that is a factor 2
lower; we adopt $2.5 \le S_{BSS} \le 5$~for the halo. The upper limit
$N_{BHB}=6$~for the bulge indicates the lower limit $N_{BSS}/N_{BHB}
\ge 3.0$. Thus our BSS fraction is marginally consistent with that of
the halo, but as yet the statistics are too poor to support further
interpretation.

In principle it would be of interest to compare the normalised blue
straggler population in the bulge with those from the dSph and the
clusters. In particular, the collection of $S_{BSS}$~vs integrated
light $M_V$~(a proxy for the mass) for clusters and dSph of
\citet{momany07} suggests differing trends for clusters and
spheroidals, against which the bulge would make an interesting
comparison. 

In practice, however, the true variation of blue straggler population
with mass, is still far from settled observationally; we use the case
of the dSph objects Draco and Ursa Minor as a case study. For these
objects, different authors reach strikingly different conclusions
using data from the same facility (the Wide Field Camera on the 2.5m
Isaac Newton Telescope). In their compilation, \citet{momany07} quote
$S_{BSS} \approx1.23, 1.35$~for Draco \citep{carrera02} and Ursa Minor
\citep{aparicio01} respectively, while \citet{mapelli07} obtain
instead $S_{BSS}\approx 0.25, 0.3$~respectively for the same
systems. (Both studies conclude that Draco and Ursa Minor are true
fossil systems, with no extensive present-day star formation
activity.)

The two studies differ in the choice of selection region for BSS on
the CMD; that of \citet{mapelli07} is about a factor two smaller than
that of \citet[][and refs therein]{momany07}, and is separated from
the MSTO by a larger distance in the CMD. Thus it is unsurprising that
the more conservative BSS region produces a lower number of BSS
candidates; our problem is choosing the study most appropriate for
comparison to our own work. The BSS selection region we have adopted
is most similar to that of \citet[][compare our Figure 1 with Figure 1
  of that paper]{momany07}, and is consistent with the selection
regions used for the cluster studies cited therein. Our belief is
therefore that \citet{momany07} is the most appropriate for comparison
to our own estimate.

In addition to photometric selection effects, the sample of objects
under consideration includes systems of differing turn-off mass, which
means the BSS populations uncovered in these systems may have quite
different formation mechanisms from each other. Furthermore, our
sample of BSS candidates is different from those of the external
systems since our candidates were found in a pencil-beam survey
through a narrow part of the bulge. A better interpretation requires
population synthesis modeling, which is beyond the scope of the
present investigation. With these caveats in mind, we present the BSS
fraction of the bulge compared to the \citet{momany07}~selection in
the hope that it will provoke just such an investigation.

\begin{figure}
\epsscale{1.0}
\plotone{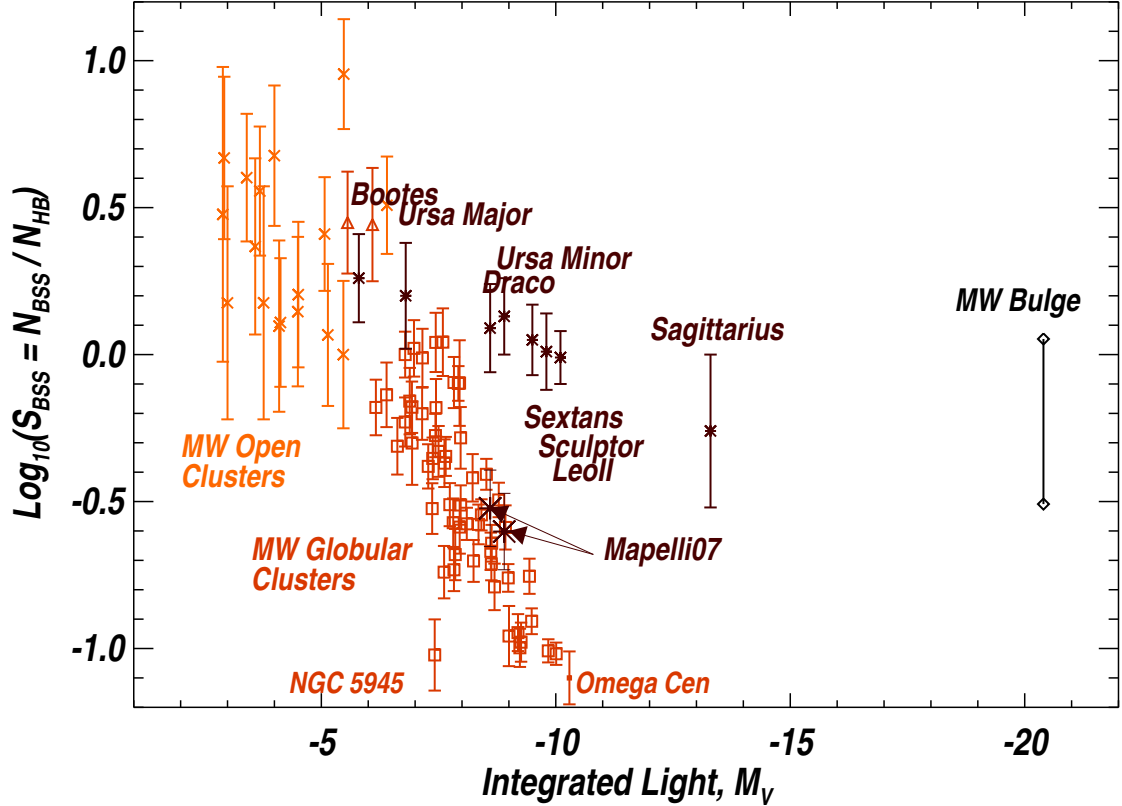}
\caption{Comparison of the normalized bulge blue straggler fraction $S_{BSS}$~(limits: black diamonds) with other stellar populations, as a function of integrated light $M_V$. These are: dwarf spheroidal galaxies without prominent current star formation (asterisks; each is labeled), Milky Way globular clusters (boxes), the Milky Way globular clusters NGC 6717 and NGC 6838, which may have significant field-star contamination (triangles) and the Milky Way open clusters (crosses). Literature points are from the compilation of Momany et al. (2007 and refs therein) except for the bulge (this work), and the redeterminations of the BSS fraction in Draco and Ursa Major by Mapelli et al. (2007; indicated with arrows as they overlap the cluster sequence). See Section \ref{s_compare}}
\label{fig_sbss}
\end{figure}

We use COBE photometry to estimate $M_V\sim 20.4$~for the bulge
\citep{dwek95}. When placed on the $S_{BSS}/M_V$~diagram, the bulge
appears to be consistent with the trend pointed out by \citet{momany07}. If
the ($M_v, S_{BSS}$)~trend reported in \citet{momany07} is indeed
borne out by further observation, then the bulge may extend this trend by
over seven magnitudes in $M_V$~(or, assuming constant M/L ratio,
nearly three orders of magnitude in mass; Figure \ref{fig_sbss}). What
this suggests about BSS evolution in the bulge as compared to the
dwarf spheroidals, is unclear at present, and we defer interpretation
until the \citet{momany07} trend has been further established or
falsified.  It should also be borne in mind that our study samples the
bulge quite differently (narrow and deep) to the dSph galaxies
(wide-area but relatively shallow); however as ours is the first study
to provide a {\it measurement} of $S_{BSS}$~for the bulge, we show its
location on the $S_{BSS}$-$M_V$~diagram to stimulate further
investigation.

We {\it can} say that the bulge is discrepant from the sequence
suggested by the globular clusters; for it to lie on the cluster trend
we would have to have observed no BSS in our sample at all.

\subsection{Young-bulge population} \label{s_young}

For the purposes of this report we define ``young objects'' as stars
already on the main sequence, with ages $\la 5$~Gy. Our photometry
covers main sequence objects up to about two solar masses, so the main
sequence lifetime of all objects in our sample is long compared to the
time spent on the Hyashi track.

The simplest estimate assumes that the fractional contribution by
young objects is the same across all luminosity bins. In this case the
young-bulge population is estimated from the number of non-BSS bulge
objects in the BSS region of the CMD (0-19; Section \ref{s_bounds})
and the total number of bulge objects in the same magnitude range and
with the same kinematic and error selection (346 objects). This yields
a young-bulge fraction in this luminosity strip, of 0-5.5\%.

A more complete estimate extrapolates the young population within the
BSS selection-box to the entire bulge sample for which we have
astrometry. This extrapolation is complicated by differing
incompleteness to astrometry above and below the MSTO (Section
\ref{s_obs}), and by our lack of knowledge of the luminosity function
of young bulge objects.

To estimate this scaling, we make two assumptions: (i). that the young
bulge population and disk population both follow the same Salpeter
mass function (i.e. $dN/dm \propto m^{-2.35}$) and the same
mass-magnitude relationship, and (ii). that the astrometric selection
effects for disk and young bulge objects vary with instrumental
magnitude in the same way. The second assumption allows us to account
for astrometric incompleteness in the young-bulge objects, while the
first allows us to correct the estimate for the fact that the disk and
bulge populations are seen at different distance moduli and thus
sample different parts of the mass function.

Denoting the number of objects within the BSS selection region as
$N_{BSS}$~and the number from the base of the BSS region in the CMD to
the faint limit of our kinematically-cleaned sample ($F814W=23$) as
$N_{faint}$~we expect
\begin{eqnarray}
  \left(\frac{N_{faint}}{N_{BSS}}\right)_{disk} & = & \left( \frac{c_{faint}}{c_{BSS}}\right)_{disk} \left(\frac{n_{faint}}{n_{BSS}}\right)_{disk} \nonumber \\
  \left(\frac{N_{faint}}{N_{BSS}}\right)_{young~bulge} & = & \left( \frac{c_{faint}}{c_{BSS}}\right)_{young~bulge} \left(\frac{n_{faint}}{n_{BSS}}\right)_{young~bulge} 
\end{eqnarray}
\label{eq_masspred}
\noindent where the quantities $c_{BSS}$~and~$n_{BSS}$~refer to the
astrometric completeness and the integral of the mass function
respectively within the magnitude limits of the BSS selection box. We
do not distinguish between the initial and present-day mass function
for the disk and young bulge populations.

Under our assumption (ii) above that
$\left(\frac{c_{faint}}{c_{BSS}}\right)_{disk} =
\left(\frac{c_{faint}}{c_{BSS}}\right)_{young~bulge}$, estimates of
$\left(\frac{n_{faint}}{n_{BSS}}\right)_{disk}$,
$\left(\frac{n_{faint}}{n_{BSS}}\right)_{young~bulge}$~and
$\left(\frac{N_{faint}}{N_{BSS}}\right)_{disk}$~allow us to solve for
$\left(\frac{N_{faint}}{N_{BSS}}\right)_{young~bulge}$. The first two
quantities may be estimated from the mass function under assumption
(i) above, while the third can be estimated from our kinematic
dataset.

We use the zero-age, solar metallicity disk isochrone of
\citet{sahu06}~to convert magnitude limits to mass limits for {\it
  both}~the disk and young-bulge populations. Investigation of a wider
range of metallicities for young objects is beyond the scope of this
report. In reality, the disk stars are distributed over a range of
distances along the line of sight, although as the largest
contribution of disk stars comes from the Sagittarius spiral arm we
restrict ourselves here to a single distance for the disk
population. The isochrone allows us to convert magnitude limits in the
``BSS'' and ``faint'' samples to mass limits for the disk population
in order to estimate $\left(\frac{n_{faint}}{n_{BSS}}\right)_{disk}
\equiv f_{MF,disk}$. To estimate this quantity for the young bulge
population, the difference in distance moduli for bulge and disk are
used to estimate the corresponding mass limits for the young-bulge
sample. Table 2 shows the mass limits adopted and the correction
factors adopted from this analysis.

\begin{figure}
\epsscale{0.6}
\plotone{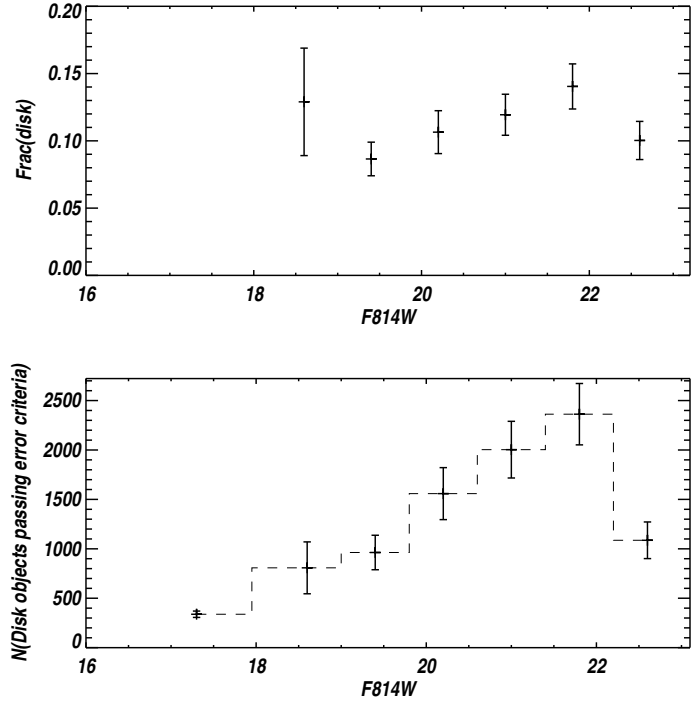}
\caption{Estimating the scaling of the number of young objects within the BSS selection region (Figure 1) to the full kinematic sample, in the presence of datasets of differing depth. For each magnitude strip, the marginal $\mu_l$~distribution is fit by a two-component gaussian, identified with the disk and bulge. The fractional contribution from the disk is then identified from the fits, and the errors estimated by Monte Carlo (Section \ref{s_kcontam}). {\it Top:} disk fraction among the magnitude strips below the BSS CMD region; the brightest bin contains the MSTO itself. {\it Bottom:} histogram of the number of disk objects detected in the sample, this time with the number of disk objects detected in the BSS region included (brightest bin). See Section \ref{s_young}.}
\label{fig_ndisk}
\end{figure}

All that remains is to estimate
$\left(\frac{N_{faint}}{N_{BSS}}\right)_{disk}$~from observation. We
follow a similar method to the fitting of the size of the bulge and
disk population within the BSS region as described in Section
\ref{s_kcontam}~above; the longitudinal proper motion distribution
$\mu_l$~is estimated for a series of magnitude strips beneath the MSTO
and the number of disk objects in each magnitude bin is estimated from
the component of the fit that corresponds to the disk population. The
results are in Figure \ref{fig_ndisk}; the number of disk objects in
the BSS region is $N_{d,BSS} = 354 \pm 23$~while the number below the
BSS region is $N_{d,faint} = 8781 \pm 617$~objects, resulting in the
scaling $\left(\frac{N_{faint}}{N_{BSS}}\right)_{disk} = 24.8 \pm
2.4$. This then means
$\left(\frac{N_{faint}}{N_{BSS}}\right)_{young~bulge} = (24.8 \pm 2.4)
\times \left(\frac{5.20}{5.84}\right) = 22.0 \pm 2.1$. We remind the
reader that we are concerned in this section entirely with the {\it
  young} bulge objects.

This means that under our ``conservative'' scenario,
the 7-19 young bulge objects suggested by our dataset within the BSS
region would scale to approximately 163-437 young bulge objects across
the entire magnitude range. As a fraction of the 12,762
kinematically-selected bulge objects, this young population would make
up as much as $1.3\% - 3.4\%$~of the bulge population, under our
``conservative'' scenario for the number of true BSS in the bulge.

The small young-bulge population is difficult to reconcile with the
recent discovery of a metallicity transition (high to low moving
outward from the galactic center) that has been reported at
$\sim$0.6-1kpc beneath the mid-plane \citep{zocc08}. If this
transition really does indicate a more metal-rich population interior
to 0.6kpc, then the lack of young bulge objects along our sight line
would suggest that this metal-rich interior population is not a
younger population. Corroborating this suggestion, \citet{ortolani95}
considered the age of the bulge in Baade's Window and found it similar
to that of the globular clusters.

In addition, a growing number of main sequence objects in the bulge
are becoming spectroscopically accessible by microlensing; at present
fifteen bulge dwarfs have been spectroscopically studied in a box
$10\degr \times 8\degr$~about the galactic center
\citep{bensby10}. Among these objects, three are spectroscopically
dated to $\la 5$~Gy, far more than the $<3\%$~we would predict. If
we have somehow vastly underestimated our kinematic contamination from
the disk, the discrepancy is only sharpened because the true sample of
young {\it bulge}~objects would then be reduced still further. At
least part of this discrepancy may be due to the fact that most of the
lensed sources are not in the inner bulge region that we probe
here. In addition, star formation rates within the bulge probably do
vary strongly over the large spatial scales probed by microlensing
studies \citep[e.g.][]{davies09}. We note that all three young objects
in the Bensby et al. (2010) sample (their objects 310, 311 and 099)
are degrees away from our field on the plane of the sky, at negative
galactic longitudes. We await further results on the spatial
distribution of metallicity and age within the bulge, with great
interest.

\section{Conclusions}\label{s_conclude}

HST observations of a low-reddening window through the bulge have
yielded the first detection of Blue Straggler Stars (BSS) in the bulge
of the Milky Way. By combining kinematic discrimination of bulge/disk
objects with variability information afforded by an $HST$~dataset of
unprecedented length (from space and for the bulge), we find that:

\noindent $\bullet$ Of the 42 BSS candidates identified with the bulge, between
18-37 are true BSS. We estimate for the horizontal branch $N_{HB} = 30$-$58$~with the same kinematic selection, suggesting $0.31 \le N_{BSS} / N_{HB} \le 1.23$. \\
$\bullet$ If the trend in normalized BSS fraction against integrated light suggested by \citet{momany07} is real, the bulge may extend this trend by nearly three orders of magnitude in mass. \\
$\bullet$ Normalized appropriately, the BSS population in the bulge is marginally consistent with that in the inner halo but the allowed range for the bulge is very broad. \\
$\bullet$ The truly-young ($\la 5$Gy) stellar population of the bulge is at most
3.4\%, but could be as low as 0 along our sight-line. \\
$\bullet$ If the recently-discovered metallicity transition $\sim 0.6-1$kpc beneath the mid-plane \citep{zocc08} does indeed indicate a population transition, then the more metal-rich population does not have a significant component of age $ \la 5$~Gy.

In the old metal rich Galactic bulge, we may therefore conclude that
blue stragglers are a small component of the stellar population 
(18-37 of 12,762 kinematically selected bulge objects), and would not
significantly affect the integrated light of similar unresolved
populations.

\section{Acknowledgments}

Based on data taken with the NASA/ESA Hubble Space Telescope obtained
at the Space Telescope Science Institute (STScI). Partial support for
this research was provided by NASA through grants GO-9750 and GO-10466
from STScI, which is operated by the Association of Universities for
Research in Astronomy (AURA), inc., under NASA contract
NAS5-26555. RMR acknowledges financial support from AST-0709479 and
also from GO-9750. MZ and DM are supported by the FONDAP Center for Astrophysics
15010003, the BASAL CATA Center for Astrophysics and Associated
Technologies PFB-06, the MILENIO Milky Way Millennium Nucleus,
P07-021F, and FONDECYT 1110393 and 1090213.

We thank Yazan al-Momany for kindly providing the cluster and dwarf
spheroidal blue straggler populations in electronic form. WIC thanks
Hans-Walter Rix, George Preston, Con Deliyannis and Christian
Howard for useful discussion, and the astronomy department at Indiana
University, Bloomington, for hospitality during early portions of this
work. We thank Ron Gilliland for his dedication of time and effort to
the production of the lightcurves on which this work is partly
based. Finally, we thank the anonymous referee, whose comments greatly
improved the presentation of this work.

{\it Facilities:} \facility{HST (ACS)}.

\clearpage

\clearpage

\clearpage

\begin{deluxetable}{llllc}
\tablewidth{0pt}
\tablecaption{Table of Bulge Blue Straggler Candidates.}
\tablehead{
\colhead{f814W} & \colhead{f606W} & \colhead{RA (J2000.0)} & \colhead{Dec (J2000.0)}
& P(d)}
\startdata
16.397 & 17.458 & 17$^h$ 58$^m$ 55.441$^s$ & -29$\degr$ 11$'$ 1.110$''$ & - \\[-2mm]
16.448 & 17.426 & 17$^h$ 59$^m$ 1.516$^s$ & -29$\degr$ 10$'$ 49.401$''$ & 1.152 \\[-2mm]
16.537 & 17.573 & 17$^h$ 58$^m$ 58.208$^s$ & -29$\degr$ 12$'$ 1.974$''$ & - \\[-2mm]
16.544 & 17.524 & 17$^h$ 59$^m$ 7.983$^s$ & -29$\degr$ 11$'$ 8.239$''$ & - \\[-2mm]
16.705 & 17.811 & 17$^h$ 59$^m$ 5.742$^s$ & -29$\degr$ 12$'$ 3.695$''$ & - \\[-2mm]
16.822 & 17.941 & 17$^h$ 59$^m$ 2.823$^s$ & -29$\degr$ 10$'$ 18.616$''$ & - \\[-2mm]
16.921 & 17.885 & 17$^h$ 59$^m$ 6.525$^s$ & -29$\degr$ 13$'$ 25.552$''$ & - \\[-2mm]
16.942 & 17.955 & 17$^h$ 58$^m$ 59.127$^s$ & -29$\degr$ 12$'$ 56.727$''$ & - \\[-2mm]
16.975 & 17.908 & 17$^h$ 59$^m$ 1.232$^s$ & -29$\degr$ 11$'$ 14.057$''$ & - \\[-2mm]
17.121 & 18.157 & 17$^h$ 59$^m$ 1.975$^s$ & -29$\degr$ 11$'$ 44.373$''$ & - \\[-2mm]
17.175 & 18.111 & 17$^h$ 59$^m$ 7.283$^s$ & -29$\degr$ 12$'$ 56.907$''$ & - \\[-2mm]
17.207 & 18.105 & 17$^h$ 58$^m$ 57.713$^s$ & -29$\degr$ 12$'$ 0.897$''$ & - \\[-2mm]
17.237 & 18.275 & 17$^h$ 59$^m$ 4.579$^s$ & -29$\degr$ 11$'$ 1.411$''$ & 8.578 \\[-2mm]
17.330 & 18.316 & 17$^h$ 59$^m$ 6.396$^s$ & -29$\degr$ 10$'$ 23.907$''$ & - \\[-2mm]
17.345 & 18.378 & 17$^h$ 59$^m$ 2.214$^s$ & -29$\degr$ 10$'$ 45.657$''$ & - \\[-2mm]
17.454 & 18.446 & 17$^h$ 59$^m$ 4.413$^s$ & -29$\degr$ 10$'$ 28.276$''$ & - \\[-2mm]
17.500 & 18.503 & 17$^h$ 58$^m$ 55.138$^s$ & -29$\degr$ 13$'$ 4.462$''$ & - \\[-2mm]
17.512 & 18.598 & 17$^h$ 59$^m$ 7.624$^s$ & -29$\degr$ 10$'$ 32.762$''$ & - \\[-2mm]
17.515 & 18.349 & 17$^h$ 58$^m$ 58.896$^s$ & -29$\degr$ 13$'$ 15.047$''$ & - \\[-2mm]
17.537 & 18.507 & 17$^h$ 59$^m$ 1.420$^s$ & -29$\degr$ 12$'$ 21.630$''$ & 1.152 \\[-2mm]
17.552 & 18.507 & 17$^h$ 58$^m$ 53.865$^s$ & -29$\degr$ 10$'$ 32.655$''$ & - \\[-2mm]
17.554 & 18.496 & 17$^h$ 58$^m$ 54.981$^s$ & -29$\degr$ 11$'$ 18.907$''$ & - \\[-2mm]
17.557 & 18.552 & 17$^h$ 59$^m$ 3.068$^s$ & -29$\degr$ 12$'$ 52.866$''$ & - \\[-2mm]
17.559 & 18.444 & 17$^h$ 59$^m$ 5.433$^s$ & -29$\degr$ 12$'$ 10.876$''$ & 0.681 \\[-2mm]
17.559 & 18.637 & 17$^h$ 58$^m$ 56.862$^s$ & -29$\degr$ 11$'$ 8.062$''$ & 4.084 \\[-2mm]
17.564 & 18.571 & 17$^h$ 58$^m$ 53.759$^s$ & -29$\degr$ 13$'$ 7.067$''$ & - \\[-2mm]
17.586 & 18.609 & 17$^h$ 59$^m$ 8.024$^s$ & -29$\degr$ 11$'$ 35.341$''$ & - \\[-2mm]
17.604 & 18.527 & 17$^h$ 59$^m$ 6.340$^s$ & -29$\degr$ 11$'$ 16.109$''$ & - \\[-2mm]
17.646 & 18.647 & 17$^h$ 59$^m$ 4.582$^s$ & -29$\degr$ 11$'$ 46.148$''$ & - \\[-2mm]
17.680 & 18.659 & 17$^h$ 58$^m$ 55.199$^s$ & -29$\degr$ 11$'$ 22.063$''$ & - \\[-2mm]
17.744 & 18.694 & 17$^h$ 58$^m$ 58.157$^s$ & -29$\degr$ 10$'$ 54.293$''$ & - \\[-2mm]
17.783 & 18.833 & 17$^h$ 58$^m$ 59.460$^s$ & -29$\degr$ 10$'$ 37.722$''$ & - \\[-2mm]
17.788 & 18.697 & 17$^h$ 59$^m$ 2.043$^s$ & -29$\degr$ 10$'$ 19.798$''$ & - \\[-2mm]
17.799 & 18.839 & 17$^h$ 59$^m$ 0.994$^s$ & -29$\degr$ 12$'$ 2.624$''$ & - \\[-2mm]
17.805 & 18.764 & 17$^h$ 59$^m$ 7.892$^s$ & -29$\degr$ 10$'$ 50.466$''$ & - \\[-2mm]
17.853 & 18.794 & 17$^h$ 59$^m$ 6.990$^s$ & -29$\degr$ 10$'$ 28.606$''$ & - \\[-2mm]
17.853 & 18.868 & 17$^h$ 58$^m$ 56.398$^s$ & -29$\degr$ 12$'$ 16.750$''$ & - \\[-2mm]
17.865 & 18.906 & 17$^h$ 59$^m$ 3.512$^s$ & -29$\degr$ 13$'$ 22.640$''$ & - \\[-2mm]
17.877 & 18.915 & 17$^h$ 59$^m$ 2.502$^s$ & -29$\degr$ 13$'$ 26.309$''$ & - \\[-2mm]
17.915 & 18.933 & 17$^h$ 58$^m$ 53.989$^s$ & -29$\degr$ 10$'$ 26.682$''$ & - \\[-2mm]
18.106 & 19.110 & 17$^h$ 59$^m$ 6.057$^s$ & -29$\degr$ 11$'$ 18.876$''$ & - \\[-2mm]
18.131 & 19.061 & 17$^h$ 58$^m$ 55.690$^s$ & -29$\degr$ 12$'$ 34.222$''$ & - 
\enddata
\label{t_properties}
\end{deluxetable}

\begin{center}
\begin{table}
\begin{tabular}{c||c|c|c}
 & Mass limits - BSS region & Mass limits - ``Faint'' region & $\left(\frac{n_{faint}}{n_{BSS}}\right)_{disk} \equiv f_{MF}$ \\
 & $(16.2 \le F814W < 18.2)$ & $(18.2 \le F814W < 23.0)$ & \\
\hline 
Disk  & $1.12 \le \frac{M}{M_{\odot}} < 1.68$ & $0.43 \le \frac{M}{M_{\odot}} < 1.12$ & $f_{MF,disk}  = 5.84$ \\
Bulge & $1.41 \le \frac{M}{M_{\odot}} < 2.11$ & $0.53 \le \frac{M}{M_{\odot}} < 1.41$ & $f_{MF,bulge} = 5.20$ \\
\hline
\end{tabular}
\caption{Mass-function correction used when scaling the number of objects within the BSS selection region to the full magnitude range $(16.2 \le F814W < 23.0$). See Section \ref{s_young}~for details.}
\end{table}
\end{center}
\label{tab_imf}
\clearpage

\end{document}